\documentclass[useAMS,usenatbib,twocolumn]{mn2e}
\usepackage{times} 
\usepackage{amsmath}
\usepackage{rotating}
\usepackage{amssymb}

\def\pasp{PASP}

\def\aap{A\&A}
\def\apj{ApJ}

\def\mnras{MNRAS}

\def\nat{Nature}

\def\apjs{ApJS}
\def\prd{Phys. Rev. D}

\newcommand{\sm}{\sc}
\newcommand{\diff}[2]{\frac{ \mathrm{d} #1}{\mathrm{d} #2}}
\newcommand{\ddiff}[2]{\frac{\mathrm{d}^2 #1}{\mathrm{d} #2^2}}
\newcommand{\partdiff}[2]{\frac{\partial #1}{\partial #2}}
\newcommand{\partddiff}[2]{\frac{\partial^2 #1}{\partial #2^2}}

\title[Caustics in growing Cold Dark Matter Haloes]
{Caustics in growing Cold Dark Matter Haloes}
\author[Vogelsberger et al.] 
{
Mark Vogelsberger$^1$\thanks{vogelsma@mpa-garching.mpg.de}, 
Simon D.M. White$^1$,
Roya Mohayaee$^2$, 
Volker Springel$^1$ \\
(1) Max-Planck Institut fuer Astrophysik,
Karl-Schwarzschild Strasse 1, 
D-85748 Garching, 
Germany \\
(2) Institut d'Astrophysique de Paris (IAP), CNRS, UPMC, 
98 bis boulevard Arago, France
}

\begin{document}
\date{Accepted ???. Received ???; in original form ???}

\pagerange{\pageref{firstpage}--\pageref{lastpage}} \pubyear{2009}

\maketitle

\label{firstpage}

\begin{abstract}
We simulate the growth of isolated dark matter haloes from self-similar and
spherically symmetric initial conditions. Our N-body code integrates the
geodesic deviation equation in order to track the streams and caustics
associated with {\it individual} simulation particles.  The radial orbit
instability causes our haloes to develop major-to-minor axis ratios
approaching 10 to 1 in their inner regions. They grow similarly in time and
have similar density profiles to the spherical similarity solution, but their
detailed structure is very different. The higher dimensionality of the orbits
causes their stream and caustic densities to drop much more rapidly than in
the similarity solution. This results in a corresponding increase in the
number of streams at each point. At 1\% of the turnaround radius
(corresponding roughly to the Sun's position in the Milky Way) we find of
order $10^6$ streams in our simulations, as compared to $10^2$ in the
similarity solution. The number of caustics in the inner halo increases by
a factor of several, because a typical orbit has six turning points rather
than one, but caustic densities drop by a much larger factor. This reduces the
caustic contribution to the annihilation radiation. For the region between 1\%
and 50\% of the turnaround radius, this is 4\% of the total in our simulated
haloes, as compared to 6.5\% in the similarity solution.  Caustics contribute
much less at smaller radii.  These numbers assume a $100~{\rm GeV~c^{-2}}$ neutralino with
present-day velocity dispersion $0.03~~{\rm cm~s^{-1}}$, but reducing the dispersion by ten
orders of magnitude only doubles the caustic luminosity.  We conclude that
caustics will be unobservable in the inner parts of haloes.  Only the
outermost caustic might potentially be detectable.
\end{abstract}


\section{Introduction}

Dark matter is thought to consist of weakly interacting particles that are
cold, meaning that their thermal velocities in unclustered regions of the
present Universe are very small.  For example, for standard values of the
relevant cross-sections, a neutralino with a mass of $100~{\rm GeV~c^{-2}}$ is
predicted to have a velocity dispersion of just $0.03$~cm/s in such regions,
corresponding to a very high phase-space density. Since the particles are
collisionless, this phase-space density is conserved along particle
trajectories, and the nonlinear collapse of dark haloes gives rise to
caustics. For infinitely cold dark matter, the density is formally infinite at
caustics, because the particles occupy a three-dimensional ``sheet'' in
six-dimensional phase-space and projection of this sheet onto configuration
space leads to singularities. These singularities are regularised in realistic
cases by the small but finite velocity dispersion of the dark matter
particles. These caustics were first discussed by Arnold et al. (1982) and
Zel'dovich et al. (1983) for the case of a neutrino-dominated universe. Their
properties were worked out in simple, one-dimensional similarity solutions for
the growth of spherical haloes by \cite{1984ApJ...281....1F} (FG hereafter) and
\cite{1985ApJS...58...39B}.

It has been suggested that caustics might have an impact on various
observables, in particular gravitational lensing
\citep[e.g.][]{1999ApJ...527...42H, 2003PhRvD..67j3502C, 2006A&A...445...43G} or annihilation radiation
\citep[e.g.][]{2001PhRvD..64f3515H,2006MNRAS.366.1217M,2007JCAP...05...15M,2008PhRvD..77d3531N, 2008MNRAS.390.1297M, 2009PhRvD..79h3526A}.
Over the last 10 years most studies of the properties of caustics have been
based on spherical models for self-similar infall such as those of FG and
\cite{1985ApJS...58...39B}, or extensions where angular momentum is introduced
to avoid purely radial orbits \citep[][]{1992ApJ...394....1W,1995PhRvL..75.2911S,2001MNRAS.325.1397N,2006PhRvD..73b3510N,2007PhRvD..75l3514N,1998PhLB..432..139S}.
As we will see below, the original similarity solutions are violently unstable
to the radial orbit instability (ROI) \citep{1973dgsc.conf..139A}.  Exact
density profiles for dark matter caustics were first calculated for the FG
model by \cite{2006MNRAS.366.1217M} (MS hereafter) following an approach
originally developed by \cite{1982PAZh....8..259Z} and
\cite{1987SvA....31..600K}.  In contrast, \cite{2001PhRvD..64f3515H} discussed
the effect of caustics on annihilation radiation using general arguments which
avoid simplified halo models, estimating that caustics might boost the
annihilation luminosity by a factor $\sim 5$ in the outer halo, a
significantly larger factor than found by MS for the spherical similarity
solution.

Although very large simulations of the formation of dark haloes are now
feasible
\citep[e.g.][]{2008Natur.454..735D,2008MNRAS.391.1685S,2009MNRAS.394..641Z},
it is still not possible to resolve caustics using standard N-body techniques.
Here we use the geodesic deviation equation (GDE) formalism presented in
\cite{2008MNRAS.385..236V} (VWHS hereafter) together with a rigorous
treatment of caustic passages derived in \cite{2009MNRAS.392..281W} (WV hereafter) to
follow the evolution of caustics in fully three-dimensional simulations of
halo formation. For simplicity, we start from self-similar, spherical initial
conditions, but the ROI causes our simulated haloes to develop into highly
elongated bars with a detailed structure quite different from that of the
spherical similarity solution.  This has substantial implications for the
number of dark matter streams predicted near the Sun and for the importance of
annihilation radiation from caustics. 

The plan of our paper is as follows. In Section 2 we describe our initial conditions 
and the numerical techniques used for
our simulations. In Section 3 we demonstrate that our N-body implementation is
working correctly by solving a simple spherical test problem. In Section 4 we
discuss results from our three-dimensional simulations, contrasting them with
the predictions of the similarity solution. We begin with the shape and the
density and velocity dispersion profiles, and we move on to fine-grained
phase-space structure and the caustic annihilation rate.  We give our
conclusions in Section 5. 

\section{Numerical techniques}

\subsection{Initial conditions}

We start our simulations from self-similar initial conditions, where most of
the particles are in the linear regime, i.e. well beyond the initial
turnaround radius.  The similarity solutions assume an Einstein-de Sitter
universe in which the {\it linear} mass perturbation $\delta M_i$ within a
sphere containing unperturbed mass $M_i$, when extrapolated to the initial
time $t_{\rm initial}$, satisfies $\delta M_i/M_i = 1.0624
(M_i/M_0)^{-\epsilon}$, where $\epsilon$ is a scaling index and $M_0$ is a
reference mass taken to be the mass within the turnaround radius at the
initial time. The equations of motion for a particle with radial distance $r$
and radial velocity $v$ can then be cast into similarity form by introducing
the variables $\lambda=r/r_{\rm ta}$ and $\tau=t/t_{\rm ta}$, where $r_{\rm
  ta} \propto M_i^{1/3 + \epsilon}$ and $t_{\rm ta} \propto M_i^{3\epsilon/2}$
are the turnaround radius and turnaround time of the particle under
consideration.  The solution $\lambda(\tau)$ of the resulting equation then
fully describes the phase-space structure of the halo at all times (see the
Appendix and FG for more details on these similarity solutions). Note that
the turnaround radius $r_{\rm ta}$ can be viewed either as a function
of enclosed mass, as here, or as a function of time, as will often be the
case in the following.

\begin{figure}
\center{
\includegraphics[width=0.4\textwidth]{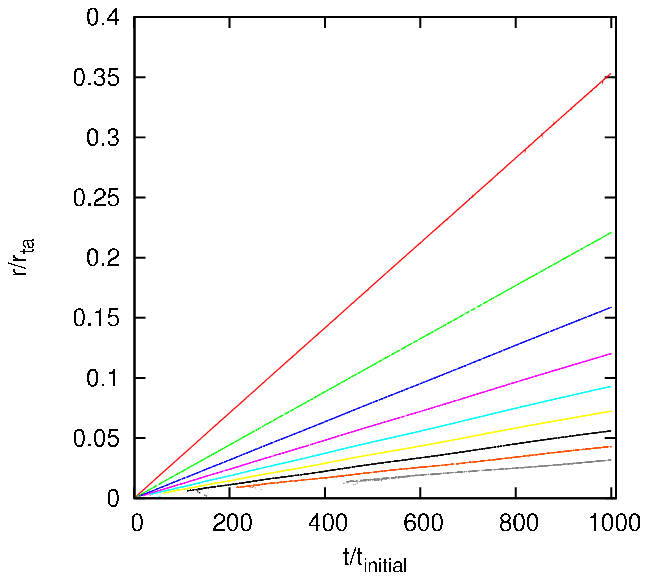}
\includegraphics[width=0.4\textwidth]{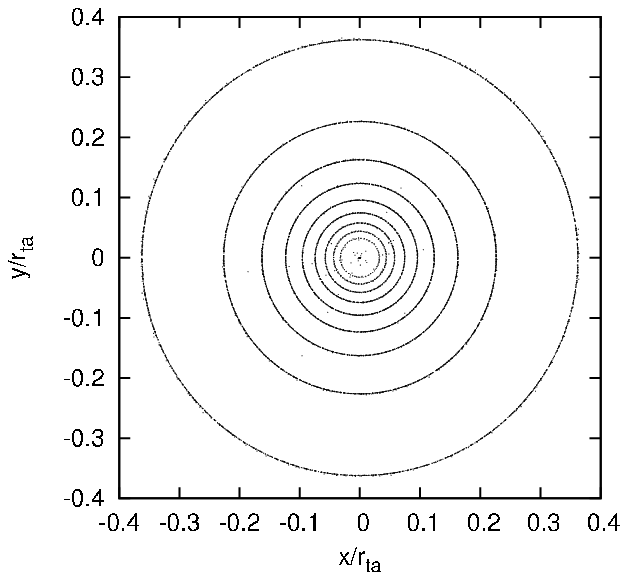}
}
\caption{Top panel: Time evolution of the caustic radii in the softened
  similarity solution ($\epsilon=2/3$, $\eta=0.15$) simulated with a
  (one-dimensional) shellcode.  Different colours represent different caustics
  from red (outermost) to grey (innermost). We note that the softened
  similarity equation only produces a finite number of caustics while the
  original FG equations produce an infinite number when approaching the halo
  centre. The sphere radii grow linearly in time because the turnaround radius
  increases linearly in time for $\epsilon=2/3$.  Bottom panel: Slice through
  the caustic spheres. The thickness of the slice is $0.0025~r_{\rm ta}(t)$. 
  We overlaid all caustics for all times using the similarity
  scaling. This produces exact spheres (rings in this slice). These plots
  demonstrate that caustics are tracked correctly over the full simulation
  period by our GDE method.}
\label{fig:CausticStructure_embedded} 
\end{figure}

We set up our initial conditions in the following way.  As a first step we
create a uniform gravitational glass within a cubic box with periodic boundary
conditions (White 1996). We then cut out the largest sphere contained within
the box and use it to represent an unperturbed Einstein-de Sitter
universe\footnote{Note that this cut means that a cubic glass with $N^3$
  particles will produce a sphere with $N^3/(6/\pi)\cong N^3/2$ particles.  In
  the following we will always specify the resolution of our simulations by
  giving the number of particles in the original cube.}.  We construct
similarity initial conditions at time $t_{\rm initial}$ by mapping this
uniform distribution $\underline{r}_{\,\rm glass}$ to the desired initial
state by scaling the radial coordinate and setting velocities according to the
similarity solution $\lambda(\tau)$
\begin{align}
 \underline{r}(t_{\rm initial}) &= r_{\rm ta}(M_0) ~\frac{\lambda(\tau)}{\tau^{2/3+2/(9\epsilon)}} ~ \frac{\underline{r}_{\,\rm glass}}{r_{\rm glass}}, \nonumber \\
 \underline{v}(t_{\rm initial}) &= \frac{r_{\rm ta}(M_0)}{t_{\rm initial}} ~\frac{{\rm d}\lambda/{\rm d}\tau}{\tau^{-1/3+2/(9\epsilon)}} ~ \frac{\underline{r}_{\,\rm glass}}{r_{\rm glass}},
\end{align}
where the similarity time variable $\tau$ is determined from the
enclosed mass according to $\tau = (M_i/M_0)^{-3\epsilon/2}$.

We run our simulations from $t_{\rm initial}$ to $t_{\rm 0}=2/(3~H_0)
\cong 9.1~{\rm Gyr}$, where we adopt $H_0=72~{\rm km}~{\rm s}^{-1}~{\rm Mpc}^{-1}
$ as the Hubble constant today.  If we require that the turnaround of the
bounding sphere occurs at $t_{\rm 0}$, choosing the scaling index $\epsilon$ 
and the mass fraction $M_0/M_{\rm tot}$ within the initial turnaround radius
determines the initial time as  $t_{\rm initial} = t_{\rm 0} (M_0/M_{\rm
  tot})^{3\epsilon/2}$ and defines a unique mapping from the uniform spherical
glass to the similarity solution at $t_{\rm initial}$. In the following we
will always use this kind of initial conditions.

\subsection{Simulation code}

For our simulations we adapt the {\sm GADGET-3} code that was developed
originally for the Aquarius project \citep{2008MNRAS.391.1685S}. We apply
vacuum boundary conditions, use (unless otherwise noted) a spline softening
kernel with constant smoothing length in comoving coordinates ($\propto
t^{2/3}$ in an Einstein-de Sitter universe), and run our simulations in
physical coordinates.  We modified the code to integrate the GDE as described
in VWHS and to track radiation from each particle due to annihilations within
its own fine-grained phase-space stream, as described in WV. This
automatically accounts for the enhanced radiation as particles pass through
caustics. We set the {\sm GADGET} force accuracy parameter to $10^{-4}$ and
chose a time integration accuracy of $10^{-3}$ \citep[see][for the
  significance of these terms]{2005MNRAS.364.1105S}.  We checked that these
settings are sufficient to produce reliable results.

\begin{figure}
\center{
\includegraphics[width=0.4\textwidth]{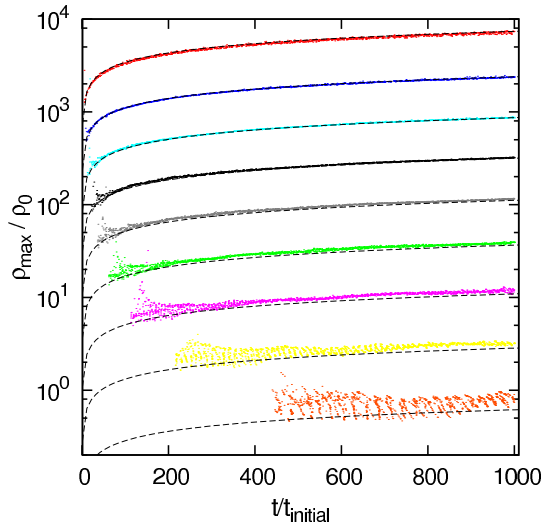}
}
\caption{Maximum caustic densities in units of the stream density at
  turnaround $\rho_0$.  Black dashed lines show maximum densities for the
  analytic solution. The densities increase with time, because the velocity
  dispersion $\sigma_{\rm b}$ decreases with time.  }
\label{fig:CausticDensity_embedded} 
\end{figure}

To gain performance and accuracy we integrate the GDE directly for each
particle only after it has passed through turnaround. At turnaround all
variables related to the fine-grained phase-space density are set to the
values predicted by the analytic FG solution. We implement this by checking in
each drift operation whether a given particle is currently turning around and
initialising the integration of the phase-space distortion tensor when this is
the case.  We demonstrate below that using the analytic solution until
turnaround is well-justified.  As a particle turns around at $t_{\rm ta}$,
its phase-space distortion tensor is set to unity and its stream density,
i.e. the local 3-density of the particular stream in which it lives, is set to
the dark matter density of the similarity solution at the turnaround point
$\rho_0 = 9 \pi^2/(16 (3 \epsilon + 1)) ~ \rho_{\mathrm{b}}(t_{\rm ta})$,
where $\rho_{\rm b}(t_{\rm ta})=1/(6\pi G t_{\rm ta}^2)$ is the mean density of the
Einstein-de Sitter background at time $t_{\rm ta}$.  The (fine-grained) dark
matter velocity dispersion $\sigma_0$ at that time and position can be
calculated using conservation of phase-space density $\rho_0/\sigma_0^3 =
\rho_{\rm b}(t_{\rm 0})/\sigma_{\rm b}(t_{\rm 0})^3$,  where $\sigma_{\rm
  b}(t_{\rm 0})$ is the dark matter velocity dispersion predicted for
unclustered matter today. For a neutralino of mass $m_p$ we have $\sigma_{\rm
  b}(t_{\rm 0}) \sim 10^{-11} {\rm c}~({\rm GeV~c^{-2}}/m_p)^{1/2}$.  Unless
otherwise noted, we will assume a $100~{\rm GeV~c^{-2}}$ neutralino, resulting
in $\sigma_{\rm b}(t_{\rm 0})=0.03~$~cm/s. Finally, the sheet orientation at
turnaround is set to
\begin{equation}
V_{q,ij}=\frac{\mathrm{d}V_i(q)}{\mathrm{d}q_j}=\frac{x_i x_j}{r^2} \left(\frac{3\pi}{4}\right)^2 \frac{1}{3+1/\epsilon} \frac{1}{t_{\rm ta}},
\end{equation}
where $V_i(q)$ are the three components of the initial cold dark matter sheet
(see VWHS and WV).  All GDE simulations presented in this paper have the
fine-grained phase-space quantities initialised for each particle at
turnaround in this way.

\section{Results}

\subsection{Embedded 1D tests}

\begin{figure}
\center{
\includegraphics[width=0.475\textwidth]{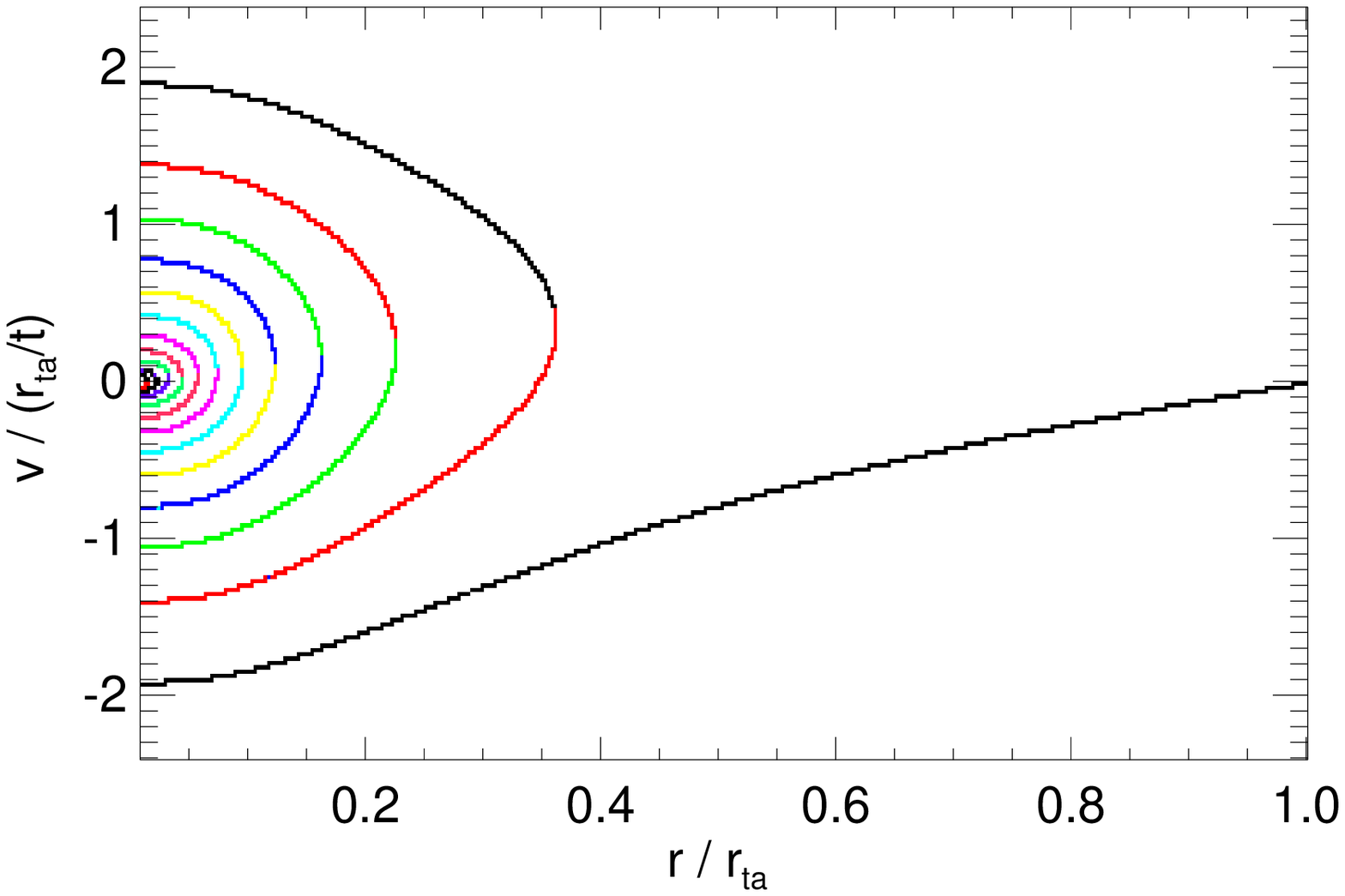}
\includegraphics[width=0.4\textwidth]{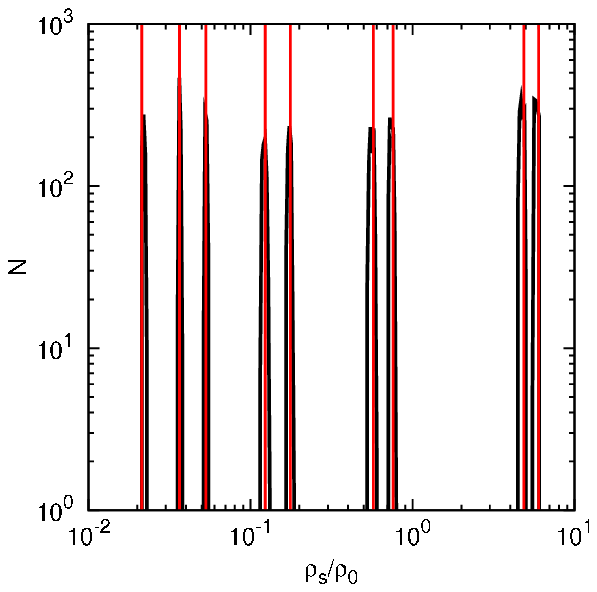}
}
\caption{Top panel: Phase-space portrait at $t_{\rm 0}$ for the softened
  similarity system.  The various colours indicate how many caustics each
  particle has passed. Infalling particles have seen no caustic (black
  line). They go through the centre, and move out to the first apocentre,
  where they pass the outermost caustic. The colour changes to red at this
  point and stays red until the second caustic is reached, and so on. We note
  that the softened similarity solution has only a finite number of
  caustics. This is why the phase-space is not densely occupied at low
  $r/r_{\rm ta}$ as in the case for the original unsoftened FG
  solution. Bottom panel: Densities of individual dark matter streams in the
  interval $r/r_{\rm ta} \in (0.1,0.105)$. At this distance the portrait shows
  nine streams (the innermost stream is coloured in yellow and particles
  belonging to it have passed four caustics).  In black we show a histogram of
  their densities in the GDE simulation and in red the analytically predicted
  stream densities. The agreement is good, demonstrating that our method can
  correctly recover the density of individual streams.}
\label{fig:Streams_embedded} 
\end{figure}
To test the GDE implementation in {\sc GADGET-3} it is desirable to have a
system where we can analytically evaluate the fine-grained phase-space
structure. This is straightforward for the FG similarity solution (see the
Appendix for details).  Starting with the initial conditions described above
we might expect to be able to reproduce the similarity solution with our
code. This is not possible, however, because the FG solution is violently
unstable to both radial and non-radial perturbations, as we show
below. Although this result does not appear to have been demonstrated before
for the specific case of the FG solution, it is expected given earlier work on
other spherical collapse models. This has found instabilities not only in the
fully 3-dimensional case where the radial orbit instability (ROI) turns the
system into a highly elongated bar \citep[e.g.][]{1973dgsc.conf..139A,
  1981SvAL....7...79P, 1985IAUS..113..297B, 1995ApJ...440....5C,
  1999PASP..111..129M, 2006ApJ...653...43M, 2008ApJ...685..739B}, but also in
1D (i.e. enforcing spherical symmetry and radial orbits) where the regular
phase-space pattern of the similarity solution still gets destroyed
\citep{1997PhRvL..78.3426H,1999MNRAS.302..321H}. These instabilities have a
dramatic impact on the fine-grained phase-space structure as we will show
below, so we cannot follow this route to check our standard code against the
known similarity solution. 

A test can nevertheless be carried out by keeping the calculation {\it
  artificially} stable.  It turns out that the following approach works. We
replaced the {\sc GADGET-3} treecode by a shellcode, where radial forces are
calculated based purely on the enclosed mass.  Particles in the initial
conditions then represent mass shells.  This removes the degree of freedom
exploited by the ROI, but does not stabilise the system against purely 1-D
instabilities. To avoid the destruction of the similarity phase-space pattern,
we must in addition soften quite strongly the gravitational potential. One can
show that a Plummer-softening that scales with the turnaround radius still
allows a similarity solution. The similarity equation is then a slight
extension to the FG equation
\begin{equation}
\frac{{\rm d}^2 \lambda}{{\rm d} \tau^2} = 
-\frac{2}{9} \left(\frac{3 \pi}{4 }\right)^2 \tau^{2/(3 \epsilon)} \frac{\lambda}{(\lambda^2 + (\eta \Lambda)^2)^{3/2}}  \mathcal{M}\left(\frac{\lambda}{\Lambda}\right),
\label{eq:softsim}
\end{equation}
where $\mathcal{M}$ is the dimensionless enclosed mass
\begin{equation}
\mathcal{M}\left(\frac{\lambda}{\Lambda}\right) =
\frac{2}{3 \epsilon} \int\limits_1^\infty \frac{{\rm d} \xi}{\xi^{1+2/(3 \epsilon)}} \mathcal{H}\left[\frac{\lambda}{\Lambda}-\frac{\lambda(\xi)}{\Lambda(\xi)}\right],
\label{eq:enclosed_mass}
\end{equation}
$\mathcal{H}$ the Heaviside function, $\Lambda(\tau) =
\tau^{2/3+2/(9\epsilon)}$ and $\eta$ is the Plummer-softening length in units
of the turnaround radius. We can solve this softened similarity equation in
the same way as the original FG equation, including the parallel GDE
integration as described in the appendix. With this setup we can check our
N-body implementation by comparing the results to the analytic solution of the
softened similarity equation.

The following results are based on a $N=128^3$ simulation with this
method. We have chosen $\epsilon=2/3$, $r_{\rm ta}(t_{\rm 0})=1400~{\rm kpc}$,
$t_{\rm initial}=t_{\rm 0}/10^3$ and $\eta=0.15$, where $r_{\rm ta}$ denotes the global
turnaround radius of the halo.

Fig.~\ref{fig:CausticStructure_embedded} shows how well our code reproduces
the caustic structure of this test problem.  The top panel shows the time
evolution of the caustic radii.  They grow linearly in time, because the
turnaround radius is proportional to $t$ for $\epsilon=2/3$.  The different
colours represent the different caustic spheres (red is the outermost sphere,
green the second, and so on). We note that the softened FG equation has only a
finite number of caustics for large $\eta$ (nine for $\eta=0.15$) while the
original FG solution (the $\eta \to 0$ limit) produces an infinite number of
caustics. The bottom panel shows a slice through the caustic spheres from a
wide range of times after scaling each to its own turnaround radius. In this
plot we thus overlay the complete time evolution of the system. One can
clearly see that the caustics build perfect spheres that scale exactly as
expected.

The maximum densities in these caustics are shown in
Fig.~\ref{fig:CausticDensity_embedded}.  The agreement of the {\sc GADGET-3}
shellcode results with the analytic solution (black dashed lines) is good. The
maximum density $\rho_{\rm max}/\rho_0$ is proportional to
$1/\sqrt{\sigma_b(t)}$ for $\epsilon=2/3$, and therefore increases with time
due to the decrease of the velocity dispersion.

In Fig.~\ref{fig:Streams_embedded} we focus on the density of individual
fine-grained dark matter streams.  We determine the density of individual
streams by binning $\rho_{s,i}/\rho_{0,i}$ for all particles $i$ in $r/r_{\rm
  ta} \in (0.1,0.105)$, where $\rho_{s,i}$ is the density of the stream
particle $i$ is embedded in and $\rho_{0,i}$ is its stream density at
turnaround.  From the phase-space portrait Fig.~\ref{fig:Streams_embedded}
(top panel) one can see that there are $9$ distinct streams in this radial
interval. In the bottom panel we show the resulting stream density histogram
in black. In red we overplot the analytic result for the densities of the nine
streams.  The agreement is good, showing that we correctly recover the density
of individual streams in this system.

We will now briefly described how we calculate the intra-stream annihilation
rate for each individual simulation particle, that is the annihilation rate due to
encounters with other particles in its own DM stream (see VWHS, WV). Each
simulation particle (mass $m_i$) represents many dark matter particles (mass
$m_p$). The intra-stream annihilation rate of one dark matter particle is
given by $(\langle \sigma_A v \rangle / m_p) \rho_{s,i}$.  Therefore, the
annihilation rate of a simulation particle is given by $(\langle \sigma_A
v \rangle / m_p^2) (\rho_{s,i} m_i)$.  To get an estimate of the intra-stream
annihilation rate, we integrate the stream density along the trajectories of
all particles as the simulation is run. This yields for every particle and at
each snapshot time $t_k$ the time-integrated rate
\begin{equation}
A_i(t_k) = \int\limits_{t_{\rm initial}}^{t_k} \mathrm{d}t ~~ \rho_{s,i}(t),
\end{equation}
where we set the particle physics prefactor $\langle \sigma_A v\rangle/
m_p^2$ to unity.  When passing through a caustic it is necessary to calculate
the time integral analytically as described in WV, since the simulation
time-stepping is usually not fine enough. Based on $A_i(t_k)$ and $A_i(t_l)$
($t_k<t_l$) we can calculate the intra-stream annihilation rate of particle
$i$
\begin{equation}
\langle P_i \rangle_{\rm intra} = m_i ~ \frac{A_i(t_l)-A_i(t_k)}{\Delta t}.
\end{equation}
In the following we apply this scheme to calculate the intra-stream
annihilation rate of individual particles.

\subsection{3D results}

\begin{figure}
\center{
\includegraphics[width=0.4\textwidth]{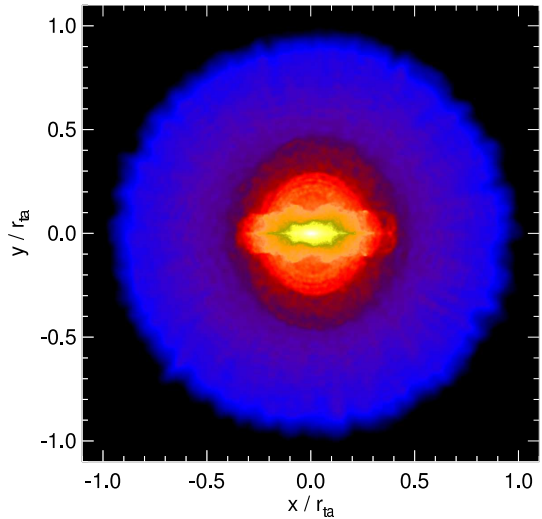}
}
\caption{Projected density of a slice of thickness $0.25 r_{\rm ta}$ through
  the 3D simulation at the final time when $r_{\rm ta}$ equals the radius of
  of the entire simulated region. The strong bar is clearly visible and is
  oriented to lie in the plane of the slice. It formed through the action of
  the ROI. The red region is bounded by the outermost caustic.}
\label{fig:density_map} 
\end{figure}
\begin{figure}
\center{
\includegraphics[width=0.4\textwidth]{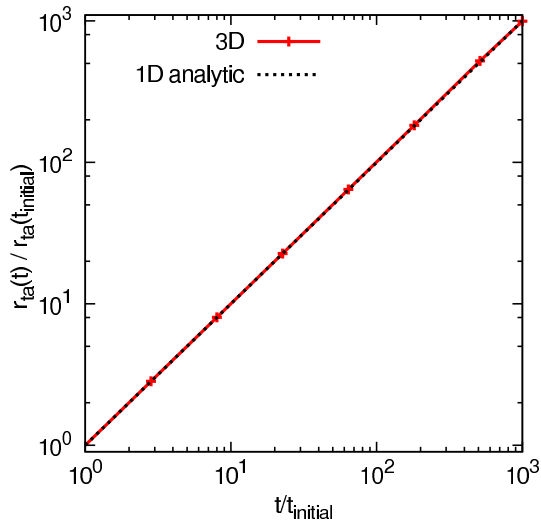}
}
\caption{The turnaround radius evolution of the 3D simulation is compared to
  the expectation from the 1D analytic similarity solution. Although the
  structure of the 3D halo is very different from that of the similarity
  solution, the turnaround radius grows in exactly the same way, linearly in
  time for $\epsilon=2/3$.}
\label{fig:rta_growth_3D} 
\end{figure}

In this section we study halo formation from the same initial conditions as
above, but with fully 3D gravity. Thus we replace the 1D shellcode of the
last section with our modified version of the tree-code GADGET-3. In addition,
we choose a much smaller softening. The ROI then acts with full force and the
system develops a highly elongated bar in its inner regions.  Our goal here is
to study how this more complicated dynamical structure affects the
fine-grained phase-space of the system, and what implications this has for the
intra-stream annihilation rate.  In the following we use a $N=128^3$
simulation with a comoving softening length of $0.01~r_{\rm ta}(t_{\rm
  0})$. We checked that our results do not depend on resolution (either
particle number or softening) in any significant way.

In Fig.~\ref{fig:density_map} we show the projected density of a slice through
the centre of the halo at $t_{\rm 0}$ with a thickness of $0.25~r_{\rm
  ta}(t_{\rm 0})$. At this time the last simulation particles are just
reaching turnaround. The blue region marks the halo beyond the outermost
caustic, hence the one-stream regime. The change to red marks the region of
multiple streams. The occupied region has a sharp spherical edge at $r_{\rm
  ta}$, showing that the influence of our vacuum boundary condition has not
propagated into the inner regions of interest.  We checked this explicitly
with other simulations, and also by verifying that the results given below are
very similar to those at $0.5 t_{\rm 0}$ once these are scaled up according to
the similarity scaling.  Clearly visible in Fig.~\ref{fig:density_map} is a
strong bar produced by the ROI. To show it optimally, we have oriented our
slice perpendicular to its minor axis.  At the time, shown we estimate axis
ratios $b/a=0.41$ and $c/a=0.22$ from the moment of inertia of all particles
within $r_{\rm ta}$, axis ratios $b/a=0.34$ and $c/a=0.17$ from all particles
within $0.5~r_{\rm ta}$, and $b/a=0.28$ and $c/a=0.12$ from those within
$0.1~r_{\rm ta}$. These numbers are quite similar at $t = 0.5 t_0$
suggesting that the system is evolving in a nearly self-similar, though
non-spherical fashion.

Although the inner structure of the halo is very different from the FG model,
the turnaround radius grows in time just as this model predicts. This is
demonstrated in Fig.~\ref{fig:rta_growth_3D}, where the dashed black line is
the FG prediction for $\epsilon=2/3$.  This behaviour is expected since the
growth of the turnaround radius depends on the monopole term in the halo mass
distribution and this is not affected by non-spherical contributions from the
inner part.  We note that this is important for our GDE integrations because
we use the analytic FG solution to initialise the GDE variables at the
turnaround radius.

\begin{figure}
\center{
\includegraphics[width=0.4\textwidth]{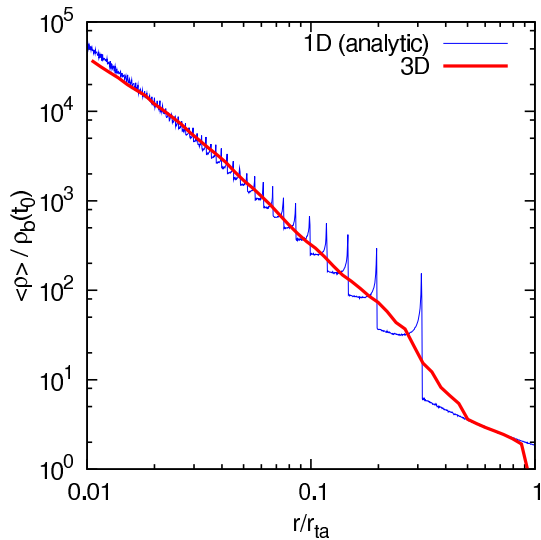}
}
\caption{Spherically averaged density profile of the 3D simulation compared to
  the FG similarity solution.  Apart from the clear caustic spikes in the 1D
  case, the two density profiles agree very well despite the very large shape
  difference between the corresponding objects. The small deviation at
  $r<0.015r_{\rm ta}$ is due to the softening of the simulation.}
\label{fig:mean_density_3D} 
\end{figure}

It is obvious that the strong departures from spherical symmetry must
nevertheless substantially change the structure of the system. The changes are
less dramatic than might be expected, however.  In
Fig.~\ref{fig:mean_density_3D} we plot the spherically averaged density
profile as a function of radius at $t_{\rm 0}$ both for the 3D simulation and
for the similarity solution. Except for the caustic spikes, the mean halo
density profile in agrees very well over most of the plotted radial range with
the similarity prediction. The deviations visible on the smallest scales in
Fig.~\ref{fig:mean_density_3D} are an effect of force softening. We have
performed simulations with 20 times smaller softening, finding that the
spherically averaged profile does not deviate significantly from isothermal
form ($\rho\propto r^{-2}$) down to $10^{-3}~r_{\rm ta}$. We have also checked
that if we change the value of the similarity exponent $\epsilon$ in our
initial conditions, the inner profiles of the resulting bar-like haloes are
always well described by a power law with exponent $9\epsilon /
(1+3\epsilon)$, the value predicted by spherical similarity solutions with
nonradial orbits
\citep{1992ApJ...394....1W,1995PhRvL..75.2911S,2001MNRAS.325.1397N}.  

The ROI disturbs the overall radial structure of the system very little. We
see no sign of a tendency to drive the system towards a ``universal'' NFW-like
profile of the kind which \cite{2006ApJ...653...43M} and
\cite{2008ApJ...685..739B} found in their own experiments from non-similarity
initial conditions. This demonstrates that the ROI does not of itself produce
NFW structure, though it may, of course, be acting in concert with other
processes in these earlier experiments.

\begin{figure}
\center{
\includegraphics[width=0.4\textwidth]{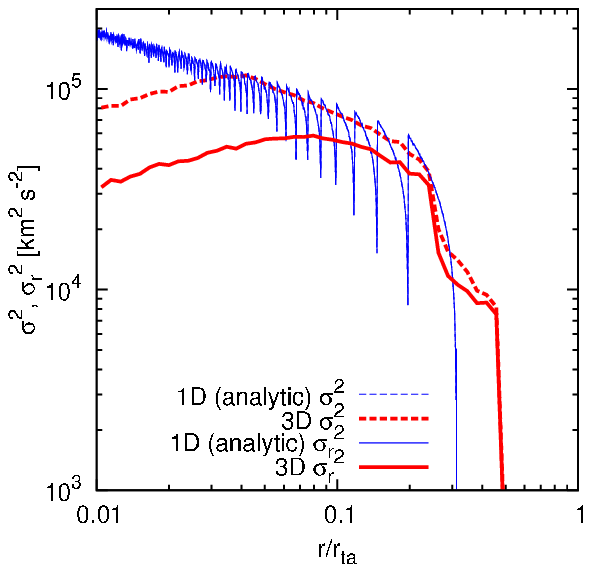}
\includegraphics[width=0.4\textwidth]{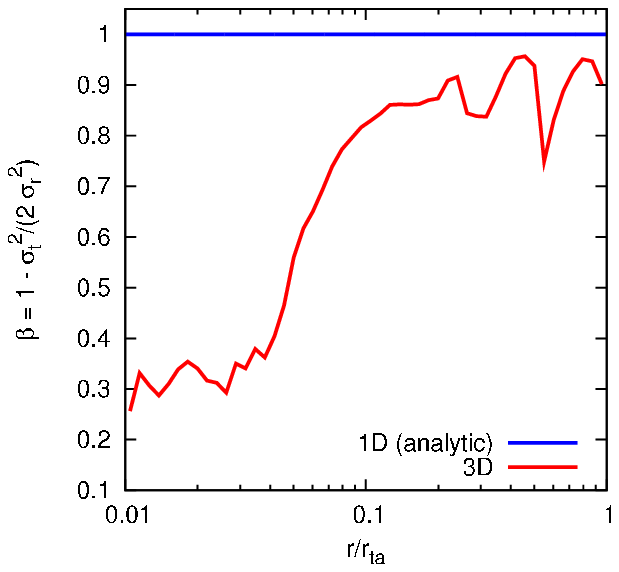}
}
\caption{Top panel: Total ($\sigma^2=\sigma_r^2+\sigma_t^2$, where $\sigma_r$
  and $\sigma_t$ are the velocity dispersions in the radial and tangential
  directions, respectively) and radial ($\sigma_r^2$) velocity dispersion
  profiles for the 3D halo are compared to the radial velocity dispersion
  profile of the FG similarity solution. Bottom panel: Velocity anisotropy
  profile $\beta=1-\sigma_t^2/(2\sigma_r^2)$.  Apart from caustic features the
  total velocity dispersion behaviour is very similar in the two models for
  $r>0.04r_{\rm ta}$. In the outer part of the halo, orbits are primarily
  radial $(\beta\sim 1)$, but the velocity distribution becomes more nearly
  isotropic for $r < 0.05 r_{\rm ta}$ and this causes both dispersions to drop
  in the inner regions as in spherical similarity solutions with nonradial
  orbits.}
\label{fig:vel_profiles_3D} 
\end{figure}

The aspherical structure of the inner regions induces non-radial motions in
the infalling material, producing a phase-space distribution which is
six-dimensional rather than two-dimensional as in the similarity solution. We
illustrate how this affects the velocity dispersion structure in
Fig.~\ref{fig:vel_profiles_3D}. The upper panel shows total and radial
velocity dispersions (relative to mean radial streaming) as functions of
radius in units of $r_{\rm ta}$, while the lower panel is a similar plot for
the velocity anisotropy parameter $\beta(r)$. In both panels the 3D simulation
result (which is at $t_{\rm 0}$) is compared with the spherical similarity
solution. The total velocity dispersion of our 3D halo tracks that in the
similarity solution quite accurately down to $0.04r_{\rm ta}$, once
the features due to the spherical caustics in the latter are smoothed
over. At smaller radii it turns over and drops significantly. This reinforces 
the conclusion from Fig.~\ref{fig:mean_density_3D} that
the coarse-grained structure of our simulated halo is similar to that of
the similarity solution despite their difference in shape. While
at $r > 0.05 r_{\rm ta}$ orbits in our bar-like halo are predominantly radial
and so resemble those in the FG similarity solution, at smaller radii the
velocity distribution becomes much closer to isotropic and the
radial velocity dispersion begins to decline towards the centre. 

\begin{figure}
\center{
\includegraphics[width=0.475\textwidth]{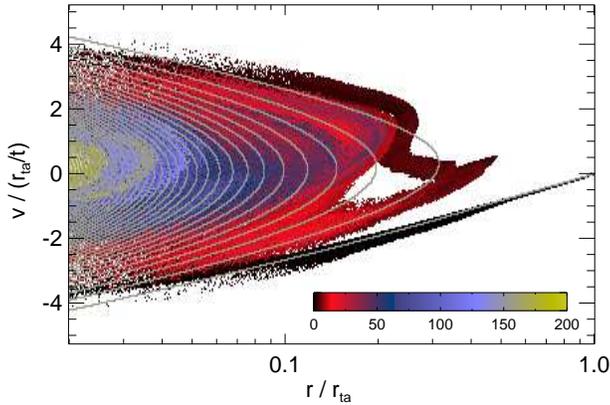}
}
\caption{Phase-space portrait at $t_{\rm 0}$ for the 3D simulation and
  (overplotted in grey) for the 1D similarity solution for $\epsilon=2/3$. The
  ROI destroys the clear phase-space pattern seen in the similarity
  solution. The coarse shape in the two cases is similar, but the 3D solution
  does not have a simple fine-grained structure in this particular projection
  of its 6-dimensional phase-space. For the 3D solution the colours encode the
  number of caustics a given particle has passed (see colourbar). One can
  clearly see that this grows towards the centre, exceeding 200 in the
  innermost regions.  }
\label{fig:phasespace_map} 
\end{figure}

The ROI leads to a considerably more complex fine-grained phase-space pattern
in our 3D simulation than in the FG similarity solution. Phase-space is
two-dimensional in the latter, with the particles occupying a one-dimensional
subspace which is fixed in similarity variables. In contrast, phase-space is
six-dimensional in our simulation, with the particles occupying a heavily
wrapped three-dimensional subspace. When this subspace is projected onto the
two-dimensional phase-space of the similarity solution, the fact that the
particles occupy a low-dimensional subspace is no longer evident. We show this
explicitly in Fig.~\ref{fig:phasespace_map}, where the pattern of the 1D
similarity solution (the grey lines) is compared to the projected particle
distribution at $t_{\rm 0}$ in our simulation. The colours in the 3D case mark
the number of caustics each particle has passed. As in the 1D case, this
number increases towards the centre, because of the shorter orbital periods at
smaller radii. Near and beyond the turnaround radius particles follow the
similarity solution, but deviations are already visible at $r\sim 0.5~r_{\rm
  ta}$, where the infalling particles no longer lie exactly on the analytic
sheet.

\begin{figure}
\center{
\includegraphics[width=0.4\textwidth]{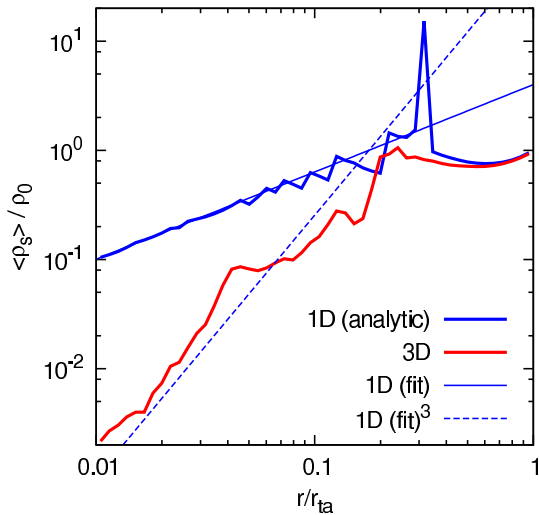}
}
\caption{Median normed stream density (i.e. the stream density surrounding
  each particle in units of its value at turnaround) for the $1$D similarity
  solution (the thick blue line) and for the fully $3$D simulation (the thick
  red line). The $1$D relation was calculated from the exact (non-softened)
  similarity solution for $\epsilon=2/3$.  Due to the fully three-dimensional
  structure of the orbits, stream densities are diluted much faster towards
  the centre in the simulation than in the similarity solution. In both cases
  the median was calculated for $50$ equal logarithmic bins in radius. This
  smooths out the caustics of the 1D solution except for the outermost few.
  The thin blue line is a power-law fit $\rho_s \propto r^{0.8}$ to the normed
  stream density in the inner part of the similarity solution. The blue dashed
  line is proportional to the 1D density dilution to the power of three and,
  as expected, agrees fairly well with the stream dilution in the inner
  regions of the 3D simulation.}
\label{fig:stream_density_normed_3D} 
\end{figure}

The increase in dimensionality of the phase-space distribution has a dramatic
effect on the density of individual dark matter streams. Away from caustics
the stream density surrounding a particular particle is expected to decrease
from its value at turnaround in proportion to $(t/t_{\rm ta})^{-1}$ for an
effectively one-dimensional system like the similarity solution, but in
proportion to $(t/t_{\rm ta})^{-3}$ for a three-dimensional system like our
simulation \citep[][ and VWHS]{1999MNRAS.307..495H} .  In
Fig.~\ref{fig:stream_density_normed_3D} we show the median of the normed
stream density (i.e. the stream density relative to its value at turnaround)
for particles in radial bins at $t = t_{\rm 0}$, again comparing results for
the simulation and for the 1D similarity solution.  Beyond the outermost
caustic, stream densities are very similar in the two cases. At smaller radii
the stream density dilution increases much faster towards the centre in the
simulation than in the similarity solution.  In the former case the inner
behaviour is very close to a power-law.  (Recall that for both models the
typical orbital period at each radius is roughly proportional to radius, so a
power law close to $r^{1}$ is expected in the 1D case.)  In the simulation
the variation of dilution with radius is less smooth, but is still moderately
well represented by a power-law which, as expected, is the cube of the one
which best fits the similarity solution.

\begin{figure}
\center{
\includegraphics[width=0.4\textwidth]{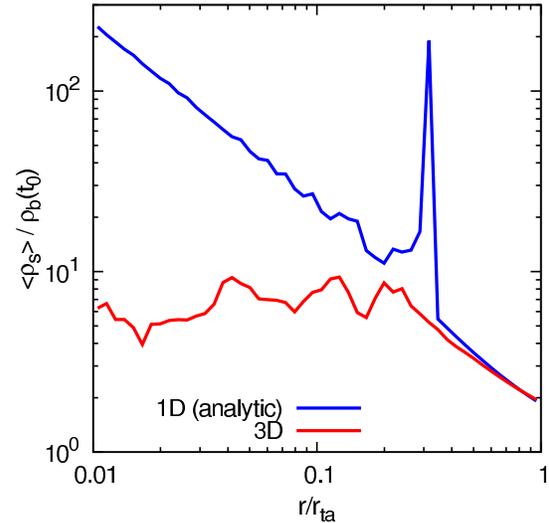}
}
\caption{The median stream density in units of the current cosmic mean density
  of the particles in each of 50 logarithmic bins in radius at time $t_{\rm
    0}$.  Results are shown for the $1$D similarity solution (blue) and for
  the $3$D simulation (red). While the stream density increases towards the
  centre for the similarity solution, it is almost constant with radius in the
  simulation.}
\label{fig:stream_density_crit_3D} 
\end{figure}

Fig.~\ref{fig:stream_density_crit_3D} presents the radial variation of stream
density in a slightly different way.  The stream density associated with each
particle is here divided by the current mean density of the Universe, rather
than by its value at turnaround. The plot is then constructed by taking
medians in 50 equal logarithmic bins in radius, just as in
Fig.~\ref{fig:stream_density_normed_3D}. In the 1D similarity solution, stream
densities increase quite strongly (approximately as $r^{-1}$) towards the
centre, because the dilution effect seen in
Fig.~\ref{fig:stream_density_normed_3D} is more than compensated by the fact
that particles near the centre typically turned around earlier, and so had
higher stream densities at turnaround. For the 3D simulation, on the other
hand, the much stronger dilution towards the centre results in typical stream
densities which are approximately constant with radius at any given time. Note
that both in this figure and in the last, medians are taken over the stream
density distribution of the particles in each radial bin. Thus they give the
stream density value that splits the {\it mass} at each radius in half. As we
will see below, this is much larger than the density of a {\it typical} stream
at that radius.

\begin{figure}
\center{
\includegraphics[width=0.4\textwidth]{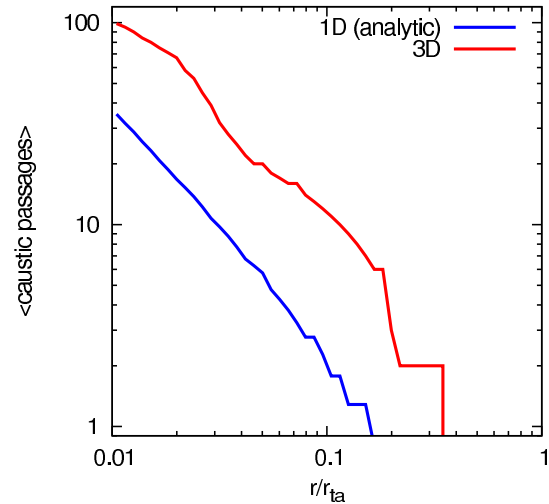}
}
\caption{Median number of caustic passages experienced by particles in a
  set of logarithmic bins in radius. The result for the 1D similarity solution
  is shown as a solid blue line, that for the 3D simulation in red.  As
  expected due to the more complex orbit structure, the number of caustics at
  a given radius increases by about a factor of three from 1D to 3D. The bump
  near $r/r_{\rm ta} =0.1$ is related to the complex feature in phase-space
  seen in Fig.~\ref{fig:phasespace_map}.}
\label{fig:caustic_passages_3D} 
\end{figure}
\begin{figure}
\center{
\includegraphics[width=0.4\textwidth]{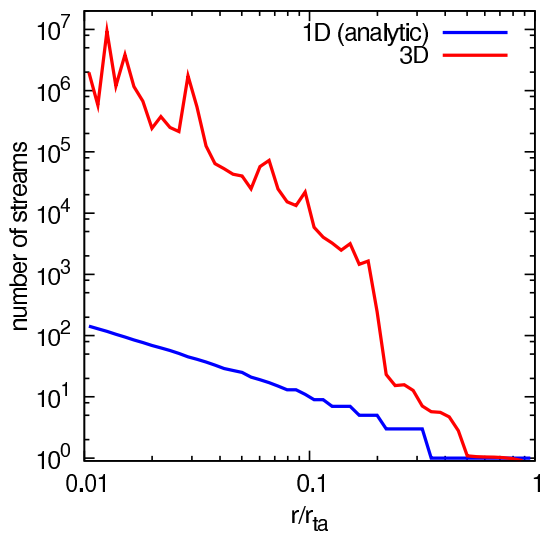}
}
\caption{Number of streams as a function of radius for the 1D similarity
  solution (blue) and the 3D simulation (red). In 3D. stream stretching reduces
  individual stream density more efficiently than in 1D, leading to a larger
  number of streams at each radius. For the 1D similarity case the number of
  streams in the FG solution is shown. In the 3D case we use the mean harmonic
  stream density of the particles in a set of logarithmic bins to estimate the
  number of streams (see text).}
\label{fig:number_of_streams_3D} 
\end{figure}
\begin{figure}
\center{
\includegraphics[width=0.4\textwidth]{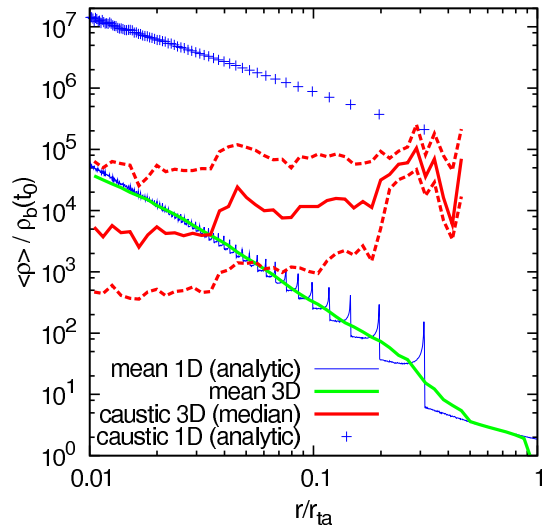}}
\caption{Densities at caustic passage as a function of radius and in units of
  the current mean cosmic density for the 1D similarity solution (blue
  crosses) and for the 3D simulation (red curves showing the median and
  quartiles at each radius).  These caustic densities are compared to the mean
  density profiles of the two systems (blue curve for the 1D similarity
  solution, green curve for the 3D simulation). In 3D the outermost caustics
  are as dense as those found in 1D, reflecting the more rapid mixing in the
  three-dimensional system.}
\label{fig:caustic_density_3D} 
\end{figure}
\begin{figure}
\center{
\includegraphics[width=0.4\textwidth]{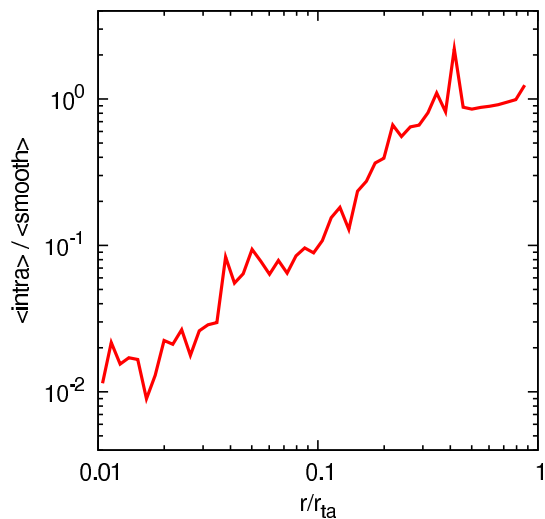}}
\caption{Local ratio of the intra-stream annihilation rate to the smoothed
  annihilation rate as a function of radius. The smoothed annihilation rate is
  calculated from an SPH-based density estimate using $64$ neighbours. As
  expected, the strongest caustic contributions are found in the outer regions.
  Near $1\%$ of the turnaround radius the contribution from caustics is at the
  percent level.  This reflects the rapid dilution of stream densities in a 3D
  system.  }
\label{fig:boostfac_3D} 
\end{figure}

Typical stream densities are quite similar near the outermost caustics in the
1D similarity solution and in the 3D simulation, but they become very
different in the inner regions. In contrast, the typical number of caustics
varies with radius in the same way in the two cases.  This can be seen in
Fig.~\ref{fig:caustic_passages_3D}, where we plot the median number of caustic
passages experienced by particles in a set of logarithmic radial bins. The
number is typically three to six times larger for the simulation than for the
similarity solution.  This is easily understood as reflecting the increased
number of turning points along typical orbits.  The overshoot in the
regions just inside the outermost caustic is related to the feature in
phase-space seen in Fig.~\ref{fig:phasespace_map}.

Since the mean density profiles are similar in the 3D and 1D cases, the lower
stream densities in 3D lead to a larger number of streams at each radius.
Counting the number of streams in 1D is straightforward, given the similarity
solution. In 3D we can estimate the number of streams crossing a given radial
bin as a suitable average over the stream densities of the particles it
contains $N_{\rm streams} \sim \rho(r) ~ \langle 1/\rho_s \rangle_r$, where
$\rho(r)$ is the mean mass density in the bin and $\langle 1/\rho_s \rangle_r$
is the average of the reciprocal of the stream densities of the particles.
The comparison is shown in Fig.~\ref{fig:number_of_streams_3D}. The greatly
enhanced mixing in 3D is clearly seen in this figure. At $1\%$ of the
turnaround radius there are of order $10^6$ streams in the 3D simulation but
only of order 100 in the FG solution. Estimates of the number of streams near
the Sun based on 1D models \citep[e.g.][]{2005PhRvD..72h3513N} give severe
underestimates of the expectation in more realistic models.

In Fig.~\ref{fig:caustic_density_3D} we show as a function of radius the
median and quartiles of the density at caustic crossing for all particles in
the 3D simulation that crossed a caustic in the $0.014~{\rm Gyr}$ immediately
preceding $t_{\rm 0}$. For comparison, we also show the mean density profile
of the halo in the 1D and 3D case. In the outer regions the density at caustic crossing is typically
a thousand times the local mean halo density. Thus one might expect caustics
to be visible in the halo annihilation signal at these radii.  In the inner
halo, however, the densities at caustic crossing do not rise as in the 1D
similarity solution (also shown for comparison) but rather stay constant or
even drop somewhat, Below $r/r_{\rm ta}=0.03$ the median density at caustic
crossing is smaller than the local mean halo density and one may expect caustics
to play no significant role in the annihilation signal.

In the inner regions, stream and caustic densities are much lower in the 3D
simulation than in the 1D similarity solution, but the number of caustics is
only a few times greater. As a result less caustic-related annihilation
radiation is expected in the 3D case.  In Fig.\ref{fig:boostfac_3D} we show
the ratio of intra-stream annihilation rate (i.e. the contribution from
particles which are both part of the same stream) to the annihilation rate
estimated from the smoothed total density. The latter is calculated by SPH
techniques using 64 neighbours. Fig.\ref{fig:boostfac_3D} shows that 
intra-stream annihilation (which is dominated by caustic annihilation) does
not contribute strongly to the overall annihilation rate. This is especially
true in the inner region of the halo, where the stream and caustic densities
are strongly suppressed. We find that for the radial interval $r/r_{\rm ta}
\in (0.01,0.5)$ the intra-stream contribution is only $4\%$. If we focus on
the outer regions we find a contribution of $24\%$ for $r/r_{\rm ta} \in
(0.1,0.5)$ and $64\%$ for $r/r_{\rm ta} \in (0.2,0.5)$, the range where the
particles turn around for the second time and produce the outermost caustics.

\begin{figure}
\center{
\includegraphics[width=0.4\textwidth]{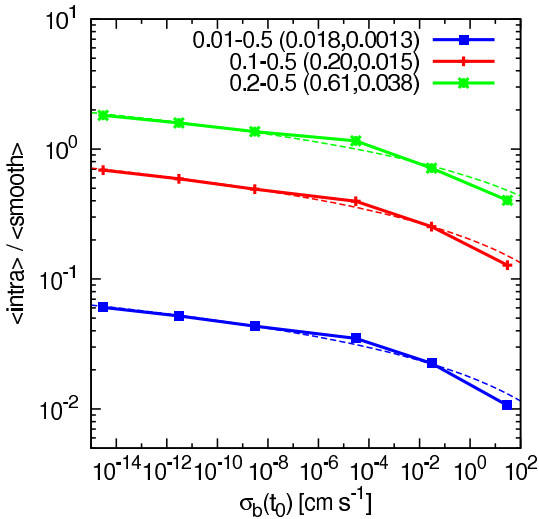}}
\caption{Intra-stream contribution to the annihilation
  rate as a function of the present-day neutralino velocity dispersion in
  unclustered regions and for different radial ranges (distinguished by
  colour; see legend which gives the range in units of the turnaround radius).
  The caustic contribution dominates and increases only logarithmically with
  decreasing dispersion. The dashed lines are analytic fits to the numerical
  results using Eq.~(\ref{eq:fit_func}). The fitting parameters $a, b$ are shown
  in the brackets in the legend.}
\label{fig:different_disp} 
\end{figure}

All the above results are based on a present-day neutralino velocity
dispersion of $\sigma_{\rm b}(t_{\rm 0})=0.03$~cm/s in unclustered regions.
The dominant contribution to the intra-stream annihilation rate comes from
caustics if the velocity dispersion is small. The densities of caustics are
proportional to $1/\sqrt{\sigma_{\rm b}}$, and so are easily scaled to an
arbitrary velocity dispersion. In Fig.~\ref{fig:different_disp} we show the
radial dependence of the intra-stream annihilation rate for a range of
velocity dispersions $\sigma_{\rm b}(t_{\rm 0})$. These results are based on
$64^3$ particle simulations with a softening length of $0.01~r_{\rm ta}$.  We
plot the intra-stream contribution for three different radial ranges as a
function of velocity dispersion. A good analytic fit to the numerical results
(dashed lines) for each radial interval is given by
\begin{equation}
\frac{\langle {\rm intra} \rangle}{\langle {\rm smooth} \rangle} = a + b
~\log\frac{1~{\rm cm/s}}{\sigma_b(t_{\rm 0})}.
\label{eq:fit_func}
\end{equation}
We note that this logarithmic behaviour is expected given the analysis
of WV and the fact that the dominant contribution comes from caustics.
From Fig.~\ref{fig:different_disp} we
conclude that decreasing the velocity dispersion by $10$ orders of magnitude
from our standard value only increases the intra-stream annihilation
contribution by a factor of about two.  In the outer part of the halo, this
would boost the total annihilation rate by almost a factor of two for the
lowest dispersion shown in the plot, but in the inner regions the caustic
contributions would still be small.  For a typical neutralino with a mass
around $100~{\rm GeV~c^{-2}}$ we do not confirm the substantial boost factors
claimed by \cite{2001PhRvD..64f3515H}.

\section{Conclusions}
We have used the geodesic deviation equation formalism recently developed by
\cite{2008MNRAS.385..236V} together with the fully general treatment of
annihilation in caustics by \cite{2009MNRAS.392..281W} to understand how
fine-grained phase-space structure affects the annihilation radiation from the
dark matter haloes that form from self-similar, spherical initial
conditions. Such objects do not evolve according to the well-known
one-dimensional similarity solutions. Rather, they turn into highly elongated
bars as a result of the radial orbit instability. These bars have mean mass
density and total velocity dispersion profiles which are very similar to those
of the relevant similarity solutions, but the loss of spherical symmetry
results in orbits that fill a 3D volume and along which the stream density
typically decreases as $1/t^3$ rather than as $1/t$ as in the similarity
solution. At any given time, typical stream densities decrease slightly
towards the centre of our simulation, whereas they increase strongly in the
similarity solution. As a consequence, there are many more streams in the
inner regions of the simulation than in the similarity solution.  At 1\% of
the turnaround radius we find $\sim 10^6$ streams in the simulation but only
about $100$ in the similarity solution. This contradicts recent claims that
the number of streams near the Sun should be relatively small
\citep[e.g.][]{2005PhRvD..72h3513N} but agrees with the estimates of
\cite{2003MNRAS.339..834H}. The number of caustics changes much less
dramatically between the two cases, with a few times more caustics in
the inner regions of the simulation than in the comparable region of the
similarity solution.  This is as expected given the higher dimensionality of
the simulation orbits.

The impact of caustics on the annihilation signal depends on their density and
number. Caustic densities in our simulation are much smaller than in the
similarity solution, but their abundance is only modestly increased. As a
result, annihilation radiation from caustics is less important in 3D than in
1D. For example, within the radial range $r/r_{\rm ta} \in (0.01,0.5)$
caustics contribute only $4\%$ of the smooth annihilation signal. If we focus
on the region containing the outermost caustics, $r/r_{\rm ta} \in (0.2,0.5)$,
this ratio is 64\%, similar to that predicted by the similarity solution.
Decreasing the velocity dispersion by $10$ orders of magnitude from our
standard value only increases the caustic contribution to the annihilation
luminosity by a factor of about two. This is because the annihilation signal
from caustic crossing depends only logarithmically on the dark matter velocity
dispersion.

Our results are based on a simplified and unrealistic halo formation model.
However, haloes growing from $\Lambda$CDM initial conditions are
expected to mix even more efficiently, because the fully three-dimensional
character of orbits is retained and small-scale structure is expected to
enhance the stretching of the phase-space sheets and hence to result in even
greater dilution of their 3-densities.  Thus, caustics will likely be less
important in the $\Lambda$CDM case than in the simple isolated halo model
discussed in the present paper. This strengthens our conclusion that 1D
similarity solutions are inadequate and misleading models for the evolution of
the fine-grained structure of real dark matter haloes. This applies not only
to the original FG solutions but also to spherically symmetric generalisations
of them
\citep[e.g.][]{1995PhRvL..75.2911S,1997PhRvD..56.1863S,2008PhRvD..78f3508D}.
These still give qualitatively incorrect predictions for the dynamical
dilution of stream densities. The inclusion of baryons would affect the
dynamics of the dark matter in the inner halo, but would neither change the
dimensionality of the orbits nor substantially modify their characteristic
timescales. Thus no qualitative changes in behaviour are expected. The
somewhat shorter orbital timescales produced by compression of the inner halo
are likely, if anything, to accelerate mixing. Our general conclusions should
thus be unaffected.

Our analysis assumes that the annihilation cross-section does not
depend on the dark matter phase-space structure. In general this
assumption is correct, but it does not hold for a recently proposed
mechanism which adopts an additional attractive force between dark
matter particles in order to increase the annihilation cross-section
through the well-known Sommerfeld enhancement process
\citep[e.g.][]{1931AnP...403..257S,2004PhRvL..92c1303H,2005PhRvD..71f3528H,2007NuPhB.787..152C,2009PhRvD..79a5014A,2009PhRvD..79h3523L}.
Away from resonances and before saturation takes place, the increase
in the annihilation cross-section scales like $1/v$, where $v$ is the
particle encounter velocity. As a result, annihilation radiation from
low-velocity-dispersion regions like subhaloes is enhanced as the
inverse of the velocity dispersion (again as long as this dispersion
exceeds the saturation level) \citep[e.g.][]{2009PhRvD..79h3539B}.
Conversely, annihilation radiation from caustics is suppressed,
because the velocity dispersion there is very high.  In such a
scenario the radiation from individual fine-grained streams is
Sommerfeld enhanced {\it away} from caustics because of their low
internal velocity dispersion there. Thus the intra-stream annihilation
is no longer dominated by caustics as it is in the standard case we
have treated in the bulk of this paper.  We will discuss this in more
detail in a future publication.

\section*{Acknowledgements}
The simulations were carried out at the
Computing Centre of the Max-Planck-Society in Garching. 
This research was supported by the DFG cluster of excellence ``Origin and
Structure of the Universe''. RM thanks French ANR OTARIE for support.
MV thanks Stephane Colombi for useful discussions.

\begin{appendix}

\section{GDE analysis of the 1D self-similar infall}
We start from the equations of motion in the similarity variables 
$\lambda = r/r_{\rm ta}$ and $\tau = t/t_{\rm ta}$ introduced by FG
($t_{\rm ta}$ and $r_{\rm ta}$ are the turnaround time and radius of 
the particle under consideration)
\begin{equation}
\frac{{\rm d}^2 \lambda}{{\rm d} \tau^2} = 
-\frac{2}{9} \left(\frac{3 \pi}{4 }\right)^2 \frac{\tau^{2/(3 \epsilon)}}{\lambda^2}  \mathcal{M}\left(\frac{\lambda}{\Lambda}\right),
\end{equation}
where $\mathcal{M}$ is the dimensionless enclosed mass (see
Eq.~(\ref{eq:enclosed_mass})), $\Lambda(\tau) = \tau^\Theta$ and $\Theta =
2/3+2/(9\epsilon)$.  The initial conditions are $\lambda(1) = 1$ and ${\rm
  d}\lambda/{\rm d} \tau (1)= 0$, corresponding to the Lagrangian physical
coordinates $q = r_{\rm ta}$ (radial distance) and $p = 0$ (radial velocity).
\begin{figure}
\center{
\includegraphics[width=0.4\textwidth]{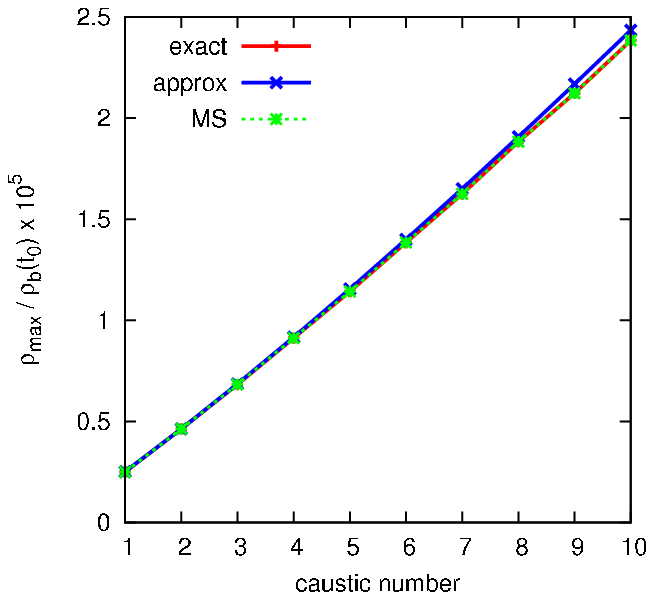}
}
\caption{Caustic density for the 1D similarity solution with $\epsilon=1$.
  Red shows the densities using the exact $\partial^2 A / \partial v^2$ values
  from the solution, while the blue line shows the result from using the
  Galilean-invariant estimate discussed in the text. The green line is the
  result obtained by MS. The agreement of our results with those presented in
  MS shows that our different approaches to calculating densities at caustic
  passage give the same result.}
\label{fig:1D_SIM_CausticDensities} 
\end{figure}
The physical distortion tensor (see VWHS) can be related to the equivalent
tensor expressed in terms of similarity variables through
\begin{align}
D_{rq} &= D_{\lambda\lambda_0},&
D_{rp} &= \frac{t}{\tau} D_{\lambda\lambda_0^\prime}, & \nonumber \\
D_{vq} &= \frac{\tau}{t} D_{\lambda^\prime\lambda_0}, &
D_{vp} &= D_{\lambda^\prime\lambda_0^\prime}. &
\end{align}
$D_{\lambda\lambda_0}$ and $D_{\lambda\lambda_0^\prime}$ are two solutions to the
the geodesic deviation equation in two-dimensional phase-space
\begin{equation}
\widetilde{D}^{\prime\prime} = 
\frac{2}{9}\left(\frac{3 \pi}{4}\right)^2 \tau^{2/(3\epsilon)} \left(\frac{2 \mathcal{M}(\lambda)}{\lambda^3} - \frac{1}{\lambda^2}\diff{\mathcal{M}(\lambda)}{\lambda}\right)\widetilde{D},
\end{equation}
\begin{figure}
\center{
\includegraphics[width=0.4\textwidth]{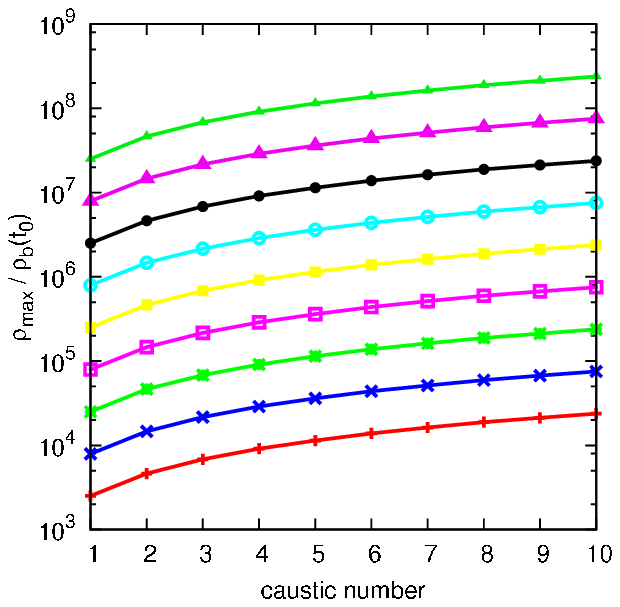}
\includegraphics[width=0.4\textwidth]{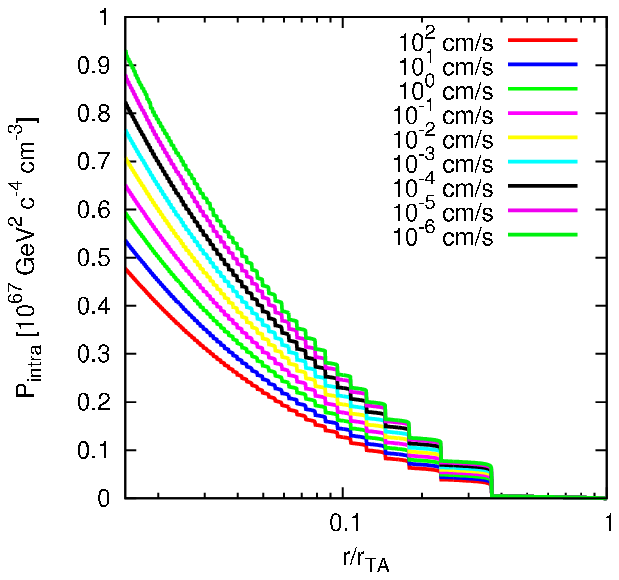}
}
\caption{Top panel: Density at caustic crossing in units of the cosmic mean
  density and for different velocity dispersions. Density scales like
  $1/\sqrt{\sigma_b}$ at each caustic. The dispersions differ by factors of 10
  so that the caustic densities differ by factors of $\sqrt{10}$. Bottom
  panel: Cumulative intra-stream annihilation rate counted outside to inside
  for the same set of dispersions. Although the dispersions change by 8 orders
  of magnitude, the intra-stream annihilation rate only changes by a factor of
  2 over the radial range shown. This is due to the logarithmic $\sigma_{\rm
    b}$ dependence of the annihilation contribution from caustic crossing.}
\label{fig:DifferentDisperions} 
\end{figure}
where $\widetilde{D}=D_{\lambda\lambda_0}$ or
$\widetilde{D}=D_{\lambda\lambda_0^\prime}$.  The initial conditions are
$\widetilde{D}(1)=1, \widetilde{D}^{\prime}(1)=0$ for $D_{\lambda\lambda_0}$
and $\widetilde{D}(1)=0, \widetilde{D}^{\prime}(1)=1$ for
$D_{\lambda\lambda_0^\prime}$. The two missing tensor components are given by
$\tau$ derivatives of $\widetilde{D}$.  To determine the stream density as a
function of $\tau$ we need the linear and second order terms
\begin{align}
\partdiff{A}{v} &= D_{\lambda\lambda_0} + \left(\frac{3 \pi}{4}\right)^2 \frac{1}{3+1/\epsilon} D_{\lambda\lambda_0^\prime}, \nonumber \\
\partddiff{A}{v} &= -t_{\rm ta} D_{\lambda\lambda_0^\prime} \,\,\, \frac{\tau^{1+\Theta}}{r_{\rm ta} \Theta^2}\left((1-\Theta)\diff{\lambda}{\tau}  +  \tau \ddiff{\lambda}{\tau}\right),
\end{align}
where $A(q,p)=p-V(q)$ and $r_{\rm ta}$ is the current turnaround radius.
The stream density for a particle is the integral of the 
phase-space density over velocity space (see WV). Using the linear and second
order results from above, this yields 
\begin{equation} 
\rho_s(\tau)\!=\!\frac{\rho_0~e^\beta K_{1/4}(\beta)}{\lambda^2\sqrt{2\pi} \sigma_0}\!\left|\partdiff{A}{v} \!\Big/\! \partddiff{A}{v}\right| , \,
\beta\!=\!\! \left(\partdiff{A}{v}\right)^{\!\!4} \!\!\Big/\! \left(2 \partddiff{A}{v}  \sigma_0 \right)^{\!\!2}, 
\end{equation} 
for the central particle of a phase-sheet, where the factor $1/\lambda^2$ is
due to the $1/(s_1 s_2)$ prefactor associated with the stretching of the
non-caustic directions and $K_{1/4}$ is the modified Bessel function.  The
intra-stream annihilation rate contribution of a small mass element
$\mathrm{d}M_i$ is given by $\mathrm{d}P_{\rm intra} =
\rho_s(\tau)~\mathrm{d}M_i$.  Integrating over all mass elements yields the
total intra-stream annihilation rate
\begin{equation}
P_{\rm intra}
= \frac{2 M_{\rm ta}}{3\epsilon} \int\limits_1^{\infty} \frac{\mathrm{d}\tau}{\tau^{1+2/(3\epsilon)}} \rho_s(\tau),
\end{equation}
where $M_{\rm ta}$ is the turnaround mass.

In Fig.~\ref{fig:1D_SIM_CausticDensities} we show the densities of the first
10 caustics calculated using this GDE approach.  For comparison to previous
work we have chosen $\epsilon=1$ and $r_{\rm ta}(t_{\rm 0})=1400~{\rm kpc}$.
We used both the exact second-order term of the similarity solution and an
estimate based on Galilean-invariant first-order quantities (see WV)
\begin{equation}
\left| \partddiff{A}{v} \right| \sim \left| \partdiff{A}{x} \right| \frac{1}{|a|} = \left| \partdiff{v}{q} \right| \frac{1}{|a|}.
\end{equation}
The green dashed line shows the results of MS. All results show good
agreement.  In Fig.~\ref{fig:DifferentDisperions} (top panel) we show the
caustic density for velocity dispersions ranging from $10^2$~cm/s down to
$10^{-6}$~cm/s. The maximum density scales as $1/\sqrt{\sigma_b}$ (see WV)
so that all lines in this plot are separated by a factor of $\sqrt{10}$.
The bottom panel shows the corresponding cumulative intra-stream
annihilation rate (outside to inside).  Although the dispersion varies over
eight orders of magnitude, the total intra-stream contribution only changes by
about a factor of $2$. This is a consequence of the logarithmic divergence of
the intra-stream annihilation luminosity near caustics (see WV).

\end{appendix}

\label{lastpage}

\end{document}